\documentclass[11pt]{article}

\usepackage{amsmath,amssymb,amsfonts}
\usepackage{graphicx}
\usepackage{bm}
\usepackage{physics}
\usepackage{hyperref}
\usepackage{geometry}
\usepackage{array}
\usepackage{authblk}
\usepackage{orcidlink}
\usepackage{hyperref}
\usepackage{amsthm}
\usepackage{xcolor}
\usepackage{tikz}
\usetikzlibrary{arrows.meta,calc}
\usetikzlibrary{positioning}
\usepackage{pdflscape}

\usepackage{cite}
\usepackage[normalem]{ulem}
\theoremstyle{remark}

\usepackage[toc,page]{appendix}

\usepackage{tabularx}
\usepackage{booktabs}
\usepackage{lineno}
\usepackage{placeins}

\title{\bf Nonlinear current dynamics and radial regularisation in the stationary Landau problem}
\author[1]{Anand Aruna Kumar\,\orcidlink{0000-0001-6148-2777}}
\affil[1]{Research Engineer, IBM Research, Albany, NY, USA\\
	\newline
	
	Email: \href{mailto:anand.aruna.kumar@ibm.com}{anand.aruna.kumar@ibm.com}}
\date{} 

\begin{document}
\maketitle
\begin{abstract}
	\noindent
	We investigate the stationary  amplitude  and current properties of a charged particle in a uniform magnetic field within the Bohm--Madelung formulation. The analysis is organised around two complementary questions: the effect of steady currents on amplitude evolution, and the influence of the singular radial behaviour on the energy spectrum of the Landau problem.
	\vspace{0.5pc}
	
    \noindent
    For the zero-current system, corresponding to the standard Landau problem, the Hamilton--Jacobi equations lead to an Ermakov--Pinney reference structure for the separated amplitudes. The azimuthal sector additionally admits periodic branches in which the magnetic flux is constrained by the angular periodicity condition. Allowing nonvanishing component currents under global current conservation introduces nonlinear amplitude evolution, most prominently through the magnetic-vector-potential coupling in the azimuthal sector. Under the minimum choice of equal and opposite radial and azimuthal currents, the radial contribution remains analytically tractable, while the nonlinear angular momentum and action are bounded through two complementary solvable comparison equations.
    \vspace{0.5pc}
	
    \noindent
    The Bohmian formulation also makes it possible to examine directly the singular structure of the radial energy equation. A Fisher-based regularisation associated with the current-branching construction is shown to closely parallel the role of Langer- and JWKB-type radial regularisation, while arising from the Bohm--Madelung amplitude--momentum description itself. The resulting modification of the radial index re-orders the Landau energy spectrum and lifts the corresponding azimuthal degeneracy.
    \vspace{0.5pc}
    
\end{abstract}

\section{Introduction}

The quantum mechanics of a charged particle in a uniform magnetic field, commonly referred to as the Landau problem, is exactly solvable in the standard Schr\"{o}dinger formulation~\cite{Landau3}. Its spectrum possesses the familiar cyclotron structure together with degeneracy associated with the angular quantum number. In the Bohm--Madelung representation,
$$
\psi=\sqrt{\rho}\,e^{iS/\hbar},
$$
the same problem is described by a Hamilton--Jacobi equation coupled to a stationary continuity equation. The formulation therefore exposes not only the energy spectrum but also the underlying relation between probability amplitude, phase momentum, and stationary probability current.

\vspace{0.5pc}

\noindent
For vanishing component currents, the stationary Bohm--Madelung equations are naturally connected with Sturm--Liouville and Ermakov--Pinney structures~\cite{AAK,Ermakov1880,Lewis1968,Reins}. The associated Ermakov--Lewis invariant formulation is well established and is used here primarily as a reference branch. The present work is concerned instead with the additional structure that emerges when stationary component currents are permitted while the total probability current remains divergence free.
\vspace{0.5pc}

\noindent
The stationary continuity equation \(\nabla\!\cdot\!\mathbf J=0\) is written in terms of component-current constants \(C_i\) satisfying
\[
C_r+C_\theta+C_z=0.
\]
We take the uniform magnetic field along the \(z\)-axis. In the symmetric gauge the vector potential has only an azimuthal component, so the magnetic coupling is confined to the transverse plane. The axial sector therefore remains free, motivating the minimal choice \(C_z=0\), for which global current conservation reduces to \(C_\theta=-C_r\).
\vspace{0.5pc}

\noindent
The two sectors respond differently. The radial amplitude retains an analytically tractable Laguerre structure, and its non-zero-current momentum can be reconstructed from
\[
\frac{1}{rR^2}\frac{d}{dr}\!\left(rR^2p_r\right)=C_r.
\]
When the radial amplitude is expanded in associated Laguerre polynomials, the accumulated current contribution reduces to finite sums of lower incomplete gamma functions. The azimuthal sector is more strongly affected by the magnetic gauge coupling. For \(C_\theta\neq0\), its momentum and amplitude become nonlinearly coupled and retain parametric dependence on \(r\). With \(C_\theta=-C_r\), the radial and azimuthal sectors are therefore linked through a common conserved-current parameter. The resulting nonlinear first-order angular equation is analysed through two complementary solvable comparison problems, which provide explicit bounds on the nonlinear momentum and the corresponding angular action.
\vspace{0.5pc}

\noindent
A second and conceptually distinct problem arises from the singular inverse-square centrifugal term in the radial Hamilton–Jacobi equation. Such singular radial terms require special treatment in semiclassical quantisation and motivate comparison with the Langer construction, which modifies the centrifugal contribution in WKB theory~\cite{Langer}. Recent exact-WKB analyses likewise emphasise the role of the regular singular point at the origin in angular-momentum-dependent quantisation~\cite{ExactWKB}. This provides a natural point of comparison with the Fisher-information regularisation previously introduced for the Bohm–Madelung equations~\cite{ASR}. In that construction, the minimum prescription is imposed on the radial canonical pair through \(rp_r=\hbar/2\), producing the shifted centrifugal factor \(l^2\rightarrow l^2+\tfrac14\) and hence the generally noninteger index
\[
\nu=\sqrt{l^2+\frac14}.
\]
The resulting radial spectrum is correspondingly reordered and the \(n_r,l\) degeneracy is lifted through a purely radial regularisation mechanism.
\vspace{0.5pc}

\noindent
The manuscript therefore develops two complementary results. First, the non-zero-current sector produces a coupled radial–azimuthal dynamics in which the radial contribution remains analytically bounded while the azimuthal evolution is controlled through nonlinear comparison solutions and their associated action bounds. Second, the radial minimum-regularisation prescription modifies the centrifugal structure and reorders the Landau spectrum by lifting the degeneracy in the azimuthal quantum number \(l\). These dynamical and spectral mechanisms are developed separately and compared within a common Bohm–Madelung framework.
\vspace{0.5pc}

\noindent
The organisation is as follows. Section~\ref{sec:sec2} briefly reviews the Ermakov reference structure used for the vanishing-current sectors. Section~\ref{sec:nonlinev} develops the stationary-current equations, beginning with the zero-current and restricted-current branches and then treating the nonlinear \(C_\theta\neq0\) azimuthal dynamics. The condition \(C_z=0\), \(C_\theta=-C_r\) is used to connect the radial and angular sectors, and the nonlinear angular equation is bounded through two complementary comparison problems. Section~\ref{sec:sec4} examines the radial inverse-square singularity from the JWKB/Langer perspective and introduces the minimum Fisher regularisation. The resulting spectral modification and degeneracy lifting are then discussed before the concluding comparison of the dynamical and regularisation mechanisms.

\section{Ermakov reference structure}
\label{sec:sec2}

Ermakov--Lewis-type invariants arise naturally in stationary
Bohm--Madelung formulations and in charged-particle dynamics in a
magnetic field~\cite{LewisRiesenfeld1969}. Their stationary and
time-dependent forms, including equivalent Riccati representations,
are discussed extensively in Refs.~\cite{Schuch1,Schuch2}. Here the
Ermakov construction is used only as the reference structure for the
vanishing-current sectors.
\vspace{0.5pc}

\noindent
For a separated Sturm--Liouville equation
\begin{equation}
	y_i''+\Omega_i^2(x_i)y_i=0,
\end{equation}
the corresponding Ermakov--Pinney amplitude satisfies
\begin{equation}
	\sigma_i''+\Omega_i^2(x_i)\sigma_i
	=
	\frac{c_i^2}{\sigma_i^3},
\end{equation}
with general solution
\begin{equation}
	\sigma_i=
	\left[
	A_i u_i^2+B_i v_i^2+2D_i u_i v_i
	\right]^{1/2},
\end{equation}
where
\begin{equation}
	A_iB_i-D_i^2=\frac{c_i^2}{W_i^2},
	\qquad
	W_i=u_iv_i'-v_iu_i'.
\end{equation}
This Wronskian construction provides the reference amplitude structure
against which the non-zero-current solutions are compared.

\subsection{Ermakov--Pinney structure and invariants in cylindrical coordinates}
\label{subsec:ermakov_explicit}
In cylindrical coordinates, the Ermakov--Pinney structure follows from the component-wise stationary continuity equations derived in ~\ref{AppendixA}. The corresponding conserved sectorial fluxes generate the nonlinear \(1/y_i^3\) terms appearing in the separated amplitude equations, consistently with the Bohm--Madelung formalism used throughout.
\vspace{0.5pc}

\subsubsection{Continuity first integrals}
For a stationary state with separable amplitude $\rho(r,\theta,z)=R^2(r)\Theta^2(\theta)Z^2(z)$,
Eq.~\eqref{eq:divfree} implies conservation of each component of the probability current,
\begin{equation}
	\frac{1}{r}\partial_r (r\rho v_r) =
	\frac{1}{r}\partial_{\theta }(\rho v_\theta) = 
	\partial_z (\rho v_z)  = 0,
	\label{eq:flux_constants}
\end{equation}
where $v_i=(\nabla S-q\mathbf A)_i/m$. Substitution of~\eqref{eq:flux_constants} into the stationary energy equation yields nonlinear amplitude equations of Ermakov--Pinney type in each sector.
\vspace{0.5pc}

\subsubsection{Canonical Sturm--Liouville and Ermakov--Pinney forms}

For each separated coordinate sector \(x_i\), we write the linear amplitude equation as
\begin{equation}
	y_i''+\Omega_i^2(x_i)\,y_i=0,
\end{equation}
and its associated Ermakov--Pinney partner as
\begin{equation}
	\sigma_i''+\Omega_i^2(x_i)\,\sigma_i=\frac{c_i^2}{\sigma_i^3}.
\end{equation}

\noindent{\em The notation $c_i$ denotes integration constants arising from~\eqref{eq:flux_constants} which we show them to be equal to ${K_i}/{\hbar}$.}
\vspace{0.1pc}

\subsubsection{Separated equations in cylindrical coordinates}
For vanishing sectorial current constants, the angular phase momentum is determined by the angular continuity first integral rather than fixed in advance by the conventional quantisation condition. We consider an amplitude ansatz of the form
\[
\sqrt{\rho(r,\theta,z)}
=R(r)\Theta(\theta;r)Z(z).
\]
The angular factor depends parametrically on the radial coordinate, so
this is not a fully separable amplitude. At fixed \(r\), however, the
angular continuity and Hamilton--Jacobi relations admit a local
one-variable reference construction.

\noindent We define the shifted angular momentum, \(P_\theta= \partial_\theta S -qBr^2/2\), which satisfies the zero current equation Eq.~\eqref{eq:flux_constants}.
\vspace{0.5pc}

\noindent
At fixed $r$, the angular continuity equation gives
\[
\frac{d}{d\theta}
\left[
\Theta^2(\theta)
\left(
\partial_\theta S-\frac{qBr^2}{2}
\right)
\right]=0.
\]
And its first integral is
\[
\Theta^2(\theta)
\left(
\partial_\theta S-\frac{qBr^2}{2}
\right) = \Theta^2(\theta)P_\theta=K_\theta(r).
\]
\vspace{0.5pc}

\noindent We introduce an angular reference parameter $l^2 $ through the angular Hamilton–Jacobi relation, we get
\begin{equation}
	P_\theta^2=\hbar^2\left(\frac{\Theta''}{\Theta}+l^2+\left(\frac{\Phi}{\Phi_0}\right)^2\right)
	\label{eq:PthetaEnergy}
\end{equation}
Here $\Phi = \pi\,r^2B$, and $\Phi_0= h/q$. 
Thus, the Ermakov--Pinney equations of the zero-current conditions are
\begin{align}
	\chi''(r)
	+ &\left[ \kappa_r^2-\frac{1}{r^2}\left(l^2 +\left(\frac{\Phi}{\Phi_0}\right)^2-\frac{1}{4}\right)
	\right ]\chi(r)
	=
	\frac{K_r^2}{\hbar^2\chi^3(r)},
	\\
	Z''(z)+ &k_z^2 Z(z)
	=
	\frac{K_z^2}{\hbar^2Z^3(z)}.
\end{align}
where \(\beta=qB/(2\hbar)\) and the azimuthal equation retains parametric dependence on $r$ due to the gauge coupling.
\vspace{0.5pc}

\noindent
The effective frequencies in the canonical Ermakov–Pinney representation are defined by
\begin{align}
	\Omega_r^2(r) &= \kappa_r^2  -\frac{1}{r^2}\left(l^2 +\left(\frac{\Phi}{\Phi_0} \right)^2-\frac{1}{4}\right), 
	\label{eq:kappar}
	\qquad \kappa^2_r=\frac{2mE}{\hbar^2}-k_z^2,\\
	\Omega_z^2(z) &= k_z^2.
\end{align}
The radial and axial sectors admit the Ermakov–Pinney reference construction. Although the angular continuity equation yields a first integral at fixed $r$, a nonconstant angular amplitude introduces additional radial derivatives into the full stationary equations. We therefore do not impose an independent angular Ermakov–Pinney reduction here; the gauge-coupled angular amplitude–momentum equations are developed in Section~\ref{sec:nonlinev}.

\subsubsection{Explicit Ermakov solutions in the separated sectors}

The effective frequencies defined above determine the corresponding
linear Sturm--Liouville equations. Two linearly independent solutions
of each equation provide the basis for constructing its
Ermakov--Pinney amplitude through the Wronskian relation.
\vspace{0.5pc}

\noindent
For the radial sector, after Liouville reduction,
\begin{equation}
	u_r''+
	\left[
	\kappa_r^2
	-\frac{1}{r^2}
	\left(
	l^2+\left(\frac{\Phi}{\Phi_0}\right)^2-\frac14
	\right)
	\right]u_r=0.
\end{equation}
Let $u_{r,1}$ and $u_{r,2}$ be two linearly independent
solutions of this equation at the same value of $\kappa_r^2$.
The corresponding Ermakov--Pinney amplitude is
\begin{equation}
	\chi_r(r)=
	\sqrt{
		A_r u_{r,1}^2(r)
		+B_r u_{r,2}^2(r)
		+2D_r u_{r,1}(r)u_{r,2}(r)
	},
\end{equation}
where
\begin{equation}
	W_r=u_{r,1}u_{r,2}'-u_{r,2}u_{r,1}',
	\qquad
	A_rB_r-D_r^2=\frac{K_r^2}{\hbar^2W_r^2}.
\end{equation}
For the standard quantum-mechanical radial branch, single-valuedness
of the angular state gives $l\in\mathbb Z$, while the radial index
depends on $|l|$. The associated-Laguerre solution therefore takes
the form~\cite{Arfken}
\begin{equation}
	u_{r,\mathrm{QM}}(r)
	\propto
	r^{|l|+\frac12}e^{-\beta r^2/2}
	L_{n_r}^{|l|}(\beta r^2),
	\qquad
	\kappa_r^2+\frac{qB|l|}{\hbar}
	=
	2\beta(2n_r+|l|+1),
\end{equation}
with $\beta>0$.
\vspace{0.5pc}

\noindent
The axial amplitude is given by
\begin{equation}
	Z(z)=
	\sqrt{
		A_z\cos^2(k_z z)
		+B_z\sin^2(k_z z)
		+2D_z\sin(k_z z)\cos(k_z z)
	},
\end{equation}
with
\begin{equation}
	W_z=k_z,
	\qquad
	A_zB_z-D_z^2=\frac{K_z^2}{\hbar^2W_z^2}.
\end{equation}
For the conventional Landau reference state, we impose
single-valuedness of the wavefunction and write
\[
\Theta=\mathrm{constant},
\qquad
\partial_\theta S=\hbar l,
\qquad
l\in\mathbb Z.
\]
The total energy is labelled $E_{\mathrm{EL}}$ and is given by
\begin{equation}
	E^{\mathrm{QM}}_{n_r l}(k_z)
	=
	\frac{\hbar\omega_c}{2}
	\left(2n_r+|l|+1\right)
	-\frac{\hbar qBl}{2m}
	+\frac{\hbar^2k_z^2}{2m},
	\qquad
	\omega_c=\frac{|q|B}{m}.
\end{equation}
\section{Stationary states with non-zero sectorial currents}
\label{sec:nonlinev}

The general form of separated continuity equations take the Liouville form
\begin{equation}
	\frac{d}{dx_i}\big(w_i u_i^2 (p_i-qA_i)\big)=C_i\, w_i u_i^2,
\end{equation}
$w_i$ are Liouville weights and $u_i$ are the respective amplitudes after Liouville scaling. Upon integration, we get

\begin{equation}
	w_i u_i^2 (p_i - qA_i)
	=
	K_i
	+
	C_i \int^{x_i} w_i u_i^2\,dx_i,
\end{equation}
with the global constraint
\begin{equation}
	\sum_i C_i = 0,
\end{equation}
ensuring overall current conservation.
\vspace{0.5pc}

\noindent In this specific problem, we solve for axial free condition, $C_z=0$ and $C_r = -C_{\theta}$ to illustrate the non-linear structure of the azimuthal component of the wave function.
\vspace{0.5pc}

\noindent
\noindent For cylindrical coordinates,
\begin{equation*}
	w_r = r, \quad w_\theta = w_z = 1.
\end{equation*}
The radial and axial sectors retain the reference structures described in Sec.~\ref{sec:sec2}. As noted there, the azimuthal amplitude retains parametric dependence on \(r\) through the magnetic gauge coupling. We therefore first construct the non-zero-current radial solution and then treat the angular dynamics locally at fixed \(r\), considering both the \(C_\theta\neq0\) nonlinear branch and the \(C_\theta=0\) reference branch without imposing the conventional Landau quantisation condition \((\partial_\theta S=l\hbar)\).

\subsection{Balanced non-zero-current branch: radial sector}
\label{sec:balanced}

The radial current equation is

$$
\frac{1}{rR^2}
\frac{d}{dr}\left(rR^2p_r\right)=C_r ,
$$
and its first integral is

$$
p_r(r)
=
\frac{
	K_r+C_r\displaystyle\int_0^r sR^2(s)\,ds
}{
	rR^2(r)
},
$$
where \(K_r\) denotes the homogeneous radial-flux contribution.
\vspace{0.5pc}

\noindent
To obtain an explicit local form, we use the associated-Laguerre radial amplitude

$$
R_{n_r\nu}(r)
=
N r^\nu e^{-\beta r^2/2}
L_{n_r}^{\nu}(\beta r^2),
$$
which provides the radial reference profile of the corresponding Landau branch. Here \(\nu\) is retained as a general radial index; in Sec.~\ref{sec:sec4} it will be shown that the minimum radial regularisation gives a generally noninteger value of \(\nu\). Writing

$$
L_{n_r}^{\nu}(x)
=
\sum_{j=0}^{n_r}a_jx^j,
\qquad
a_j
=
(-1)^j
\frac{\Gamma(n_r+\nu+1)}
{\Gamma(\nu+j+1)(n_r-j)!j!},
$$
the accumulated radial term becomes

$$
\int_0^{\beta r^2}
x^\nu e^{-x}
\left[L_{n_r}^{\nu}(x)\right]^2dx
=
\sum_{j,k=0}^{n_r}
a_ja_k\,
\gamma\!\left(
\nu+j+k+1,\beta r^2
\right),
$$
where \(\gamma(a,x)\) is the lower incomplete Gamma function ~\cite{DLMF}.
\vspace{0.5pc}

\noindent
For \(K_r=0\), the corresponding radial momentum is therefore
\begin{equation}
	p_r(r)
	=
	\frac{C_r}{2\beta^{\nu+1}}
	\frac{
		\displaystyle\sum_{j,k=0}^{n_r}
		a_ja_k\,
		\gamma\!\left(
		\nu+j+k+1,\beta r^2
		\right)
	}{
		r^{2\nu+1}
		e^{-\beta r^2}
		\left[L_{n_r}^{\nu}(\beta r^2)\right]^2
	}.
\end{equation}
The normalisation constant \(N\) cancels from this expression. The result gives the radial momentum associated locally with the chosen non-zero-current branch; its compatibility with the complete stationary Hamilton--Jacobi system is considered separately from the current integration itself.
\vspace{0.5pc}

\subsection{Balanced non-zero-current branch: azimuthal sector}
\noindent The azimuthal component is therefore of primary interest in this work, as it exhibits nonlinear amplitude-momentum coupling with radial parametric dependence. The stationary current and energy equations for the azimuthal sector are
\begin{equation}
	\label{eq:angulargaugecomp}
	\frac{d p_\theta}{d\theta}
	+\left(p_\theta-\frac{qBr}{2}\right)\frac{d}{d\theta}\ln \Theta^2
	=
	rC_\theta,
\end{equation}
\begin{equation}
	\label{eq:angularenergycomp}
	\frac{1}{2m}\left(p_\theta^2-qBr\,p_\theta\right)
	-\frac{\hbar^2}{2mr^2}\frac{\Theta''}{\Theta}
	=
	\frac{\hbar^2l^2}{2mr^2}.
\end{equation}

\noindent Using the Hamilton--Jacobi representation $p_\theta = \partial_\theta S / r$, the gauge-induced magnetic field ({$B$})-dependence becomes explicit in the azimuthal dynamics.
\vspace{0.5pc}

\noindent	A useful reduction is obtained by introducing the logarithmic derivative
\begin{equation}
	w(\theta):=\frac{\Theta'(\theta)}{\Theta(\theta)}.
\end{equation}
Then
\begin{equation}
	\frac{\Theta''}{\Theta}=w'+w^2.
\end{equation}
Using the shifted angular momentum variable
\begin{equation}
	P_{\theta}(\theta):=\partial_\theta S-\frac{qBr^2}{2},
\end{equation}
the local angular system becomes
\begin{equation}
	P_{\theta}'+2wP_{\theta}=r^2C_\theta,
	\label{eq:u-w-current}
\end{equation}
and from the energy term, 
\begin{equation}
	P_\theta^2
	=
	\hbar^2(w'+w^2+\Lambda_\theta),
	\qquad
	\Lambda_\theta
	:=
	l^2+\left(\frac{\Phi}{\Phi_0}\right)^2.
	\label{eq:u-w-energy}
\end{equation}
\noindent Eqs.~\eqref{eq:u-w-current} and ~\eqref{eq:u-w-energy} form the coupled first-order nonlinear system ~\eqref{eq:uw-system}.

\begin{equation}
	\begin{aligned}
		P_\theta' &= r^2C_\theta-2wP_\theta,\\
		w' &= \frac{P_\theta^2}{\hbar^2}-\Lambda_\theta-w^2.
	\end{aligned}
	\label{eq:uw-system}
\end{equation}
Eliminating $w(\theta)$, one obtains a second-order nonlinear differential equation for $P_\theta(\theta)$, given by

\begin{equation}
		3P_\theta'^2
		-2P_\theta P_\theta''
		-4r^2C_\theta P_\theta'
		-\frac{4}{\hbar^2}P_\theta^4\\
		+{4\Lambda_\theta}P_\theta^2
		+r^4C_\theta^2=0.
	\label{eq:nonlinpie}
\end{equation}
\noindent The above differential equation is a nonlinear equation of $P_\theta(\theta)$ and a zero-current case $C_\theta = 0$ admits a reduction to a first-order linear equation for $\xi^2(P_\theta)$. In the next subsections we use this condition to obtain closed form expression for the azimuthal amplitude wave function, $\Theta(\theta)$.

\subsection{Reference angular solution and current-driven extension}

\subsubsection{$C_\theta=0$ reference solution}

Setting $C_\theta=0$ in Eq.~\ref{eq:nonlinpie}, introduce
$\xi=P_\theta'/P_\theta$. The resulting equation
reduces to a first-order linear equation for
$\xi^2(P_\theta)$, giving
\begin{equation}
	\frac{P_\theta'}{P_\theta}
	=\pm\sqrt{
		\frac{\mathcal {E_\theta}}{\hbar} P_\theta
		-\frac{4P_\theta^2}{\hbar^2}
		-4\Lambda_\theta
	},
\end{equation}
where $\mathcal E_\theta$ is an integration constant.
The discriminant of the quadratic expression is
\begin{equation}
	\Delta_\theta
	=\mathcal E_\theta^2
	-\frac{64\Lambda_\theta}{\hbar^2}.
\end{equation}
For $\Lambda_\theta>0$ and $\Delta_\theta>0$, integration
gives the reference shifted angular momentum
\begin{equation}
	P_0(\theta;r)
	=
	\frac{8\hbar\Lambda_\theta}{
		\mathcal E_\theta+
		\sqrt{\Delta_\theta}
		\sin\!\left[
		2\sqrt{\Lambda_\theta}
		(\theta-\theta_0)
		\right]
	},
\end{equation}
on an interval where the denominator is non-zero.
\vspace{0.5pc}

\noindent
The reference angular action follows from
\[\partial_\theta S_{\theta,0}
=P_0+\hbar\Phi/\Phi_0.\]
Using the discriminant $The$ defined above,
integration gives

\begin{equation}
	\begin{aligned}
		S_{\theta,0}(\theta)
		={}S_{\theta,0}(\theta_0)
		+\hbar\frac{\Phi}{\Phi_0}(\theta-\theta_0)
		+\hbar\tan^{-1}\!\left[
		\frac{8\sqrt{\Lambda_\theta}\sin (\sqrt{\Lambda_\theta}(\theta-\theta_0))}
		{\left[\mathcal E_\theta \cos (\sqrt{\Lambda_\theta}(\theta-\theta_0))
			+\sqrt{\Delta_\theta}\sin (\sqrt{\Lambda_\theta}(\theta-\theta_0))\right]}
		\right].
	\end{aligned}
\end{equation}

\noindent
The reference momentum and action retain parametric
dependence on $r$ through the magnetic flux and
$\Lambda_\theta$. This dependence arises within the
reduced angular construction; compatibility with the
complete two-variable stationary system requires
separate examination.
\vspace{0.5pc}

\noindent
The bounded angular-momentum modulation and its associated action retain parametric dependence on $r$ through the magnetic flux. A nonconstant angular amplitude may consequently introduce additional radial-derivative coupling in the full stationary equations; its treatment is beyond the scope of the present work.
\vspace{0.5pc}

\noindent
From the zero-current first integral
\begin{equation}
	\Theta_0^2(\theta;r)\,P_0(\theta;r)=K_0(r),
\end{equation}
the angular amplitude follows directly as
\begin{equation}
	\Theta_0^2(\theta;r)
	=
	\frac{K_0(r)}{8\hbar\Lambda_\theta}
	\left[
	\mathcal E_\theta
	+
	\sqrt{\Delta_\theta}
	\sin\!\left(
	2\sqrt{\Lambda_\theta}(\theta-\theta_0)
	\right)
	\right].
\end{equation}

\noindent
Periodic boundary conditions,
\[
\Theta_0(\theta+2\pi;r)=\Theta_0(\theta;r),
\]
therefore require
\[
2\sqrt{\Lambda_\theta}=N,
\qquad N=1,2,3,\ldots,
\]
so that
\begin{equation}
	\Lambda_\theta=\frac{N^2}{4}.
\end{equation}

\noindent
Using
\begin{equation}
	\Delta_\theta=\mathcal E_\theta^2-64\Lambda_\theta,
\end{equation}
the real reference branch satisfies
\begin{equation}
	\mathcal E_\theta\geq4N,
\end{equation}
with $\mathcal E_\theta=4N$ corresponding to the constant-amplitude limit and
$\mathcal E_\theta>4N$ to the nonconstant modulated branch. 
\vspace{0.5pc}

\noindent
Furthermore,
\begin{equation}
	\int_{\theta_0}^{\theta_0+2\pi}
	P_0(\theta;r)\,d\theta
	=
	\pi N\hbar.
\end{equation}

\noindent Since
\[
\Lambda_\theta
=
l^2+\left(\frac{\Phi}{\Phi_0}\right)^2,
\]
the periodic branch also satisfies
\begin{equation}
	\left|\frac{\Phi}{\Phi_0}\right|
	=
	\sqrt{\frac{N^2}{4}-l^2},
	\qquad
	N^2\geq4l^2.
\end{equation}
\begin{figure}[!h]
	\centering
	\includegraphics[
	width=0.95\textwidth,
	height=.57\textheight,
	keepaspectratio
	]{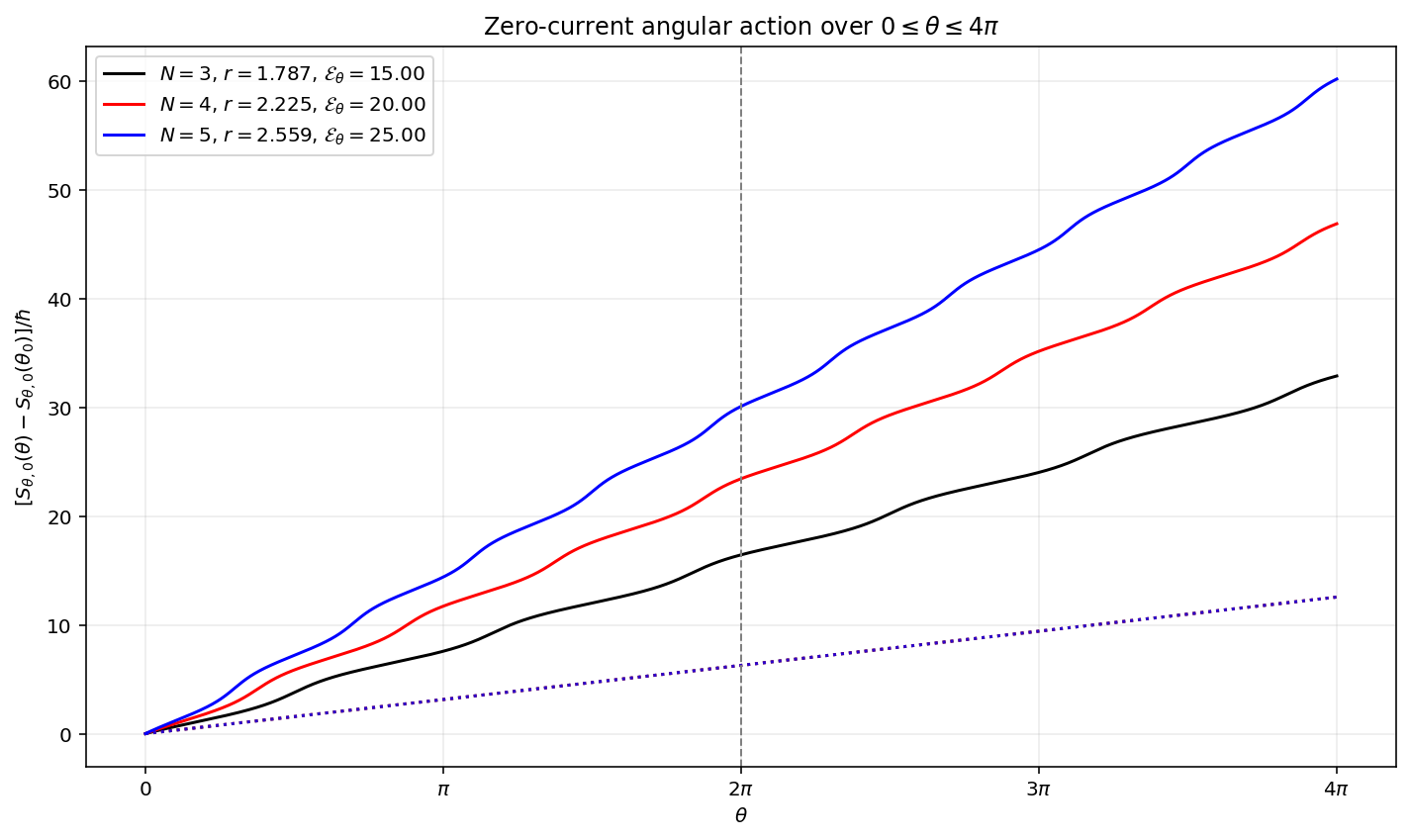}
	\caption{Zero-current angular action \(S_{\theta,0}(\theta)\) over two angular periods for the periodic branches \(N=3,4,5\). The solid curves show the continuously unwrapped action obtained from the nonlinear reference momentum, with \(\mathcal E_\theta>4N\), while the dotted straight line shows the conventional linear phase \(S_\theta^{\rm QM}=\hbar l\theta\) for comparison. Increasing \(N\), and hence the associated flux parameter, increases the mean phase slope while preserving a bounded smooth modulation about the quasi-linear evolution.}
	\label{fig:SthetaCzero}
\end{figure}
\FloatBarrier
\vspace{0.5pc}

\noindent
Figure~\ref{fig:SthetaCzero} illustrates these periodic zero-current branches through the corresponding angular action, showing how the bounded modulation is superposed on the quasi-linear phase accumulation for different admissible values of \(N\).
\vspace{0.5pc}

\noindent
The figure shows that increasing \(N\) changes the mean phase slope and modulation while preserving the same qualitative staircase-like action profile over successive angular periods.
\vspace{0.5pc}

\noindent
The periodic zero-current branch therefore combines discrete angular periodicity with a bounded, flux-dependent modulation of the otherwise quasi-linear angular action.

\subsection{Non-zero current and bounds on the angular action}
\label{sec:34}

For non-zero current, the separated radial and angular relations may be
written as
\begin{equation}
	w_i u_i^2 (p_i-qA_i)
	=
	K_i
	+
	C_i\int^{x_i}w_i u_i^2\,dx_i,
	\qquad i=r,\theta .
\end{equation}
Solving separately for the radial and angular momenta gives
\begin{align}
	p_r
	&=
	qA_r
	+
	\frac{K_r}{w_r u_r^2}
	+
	\frac{C_r}{w_r u_r^2}
	\int^{r}w_r u_r^2\,dr ,
	\\
	p_\theta
	&=
	qA_\theta
	+
	\frac{K_\theta}{w_\theta u_\theta^2}
	+
	\frac{C_\theta}{w_\theta u_\theta^2}
	\int^{\theta}w_\theta u_\theta^2\,d\theta .
\end{align}
The corresponding actions are therefore
\begin{align}
	S_r(r)
	&=
	\int^r p_r(\tilde r)\,d\tilde r ,
	\\
	S_\theta(\theta)
	&=
	\int^\theta p_\theta(\tilde\theta)\,d\tilde\theta .
\end{align}
The radial and angular currents satisfy
\begin{equation}
	C_\theta=-C_r .
\end{equation}
Over the corresponding intervals, the action increments may be collected
in the form
\begin{equation}
	\Delta S_r=U_r+C_r V_r,
	\qquad
	\Delta S_\theta=U_\theta+C_\theta V_\theta .
\end{equation}
Imposing the compatibility condition
\begin{equation}
	\Delta S_r+\Delta S_\theta=0
\end{equation}
then gives
\begin{equation}
	U_r+U_\theta+C_r(V_r-V_\theta)=0 ,
\end{equation}
and therefore
\begin{equation}
	\boxed{
		C_r=
		\frac{U_r+U_\theta}{V_\theta-V_r}},
	\qquad
	C_\theta=-C_r .
\end{equation}
The non-zero-current angular dynamics are governed by the nonlinear
first-order equation obtained in Eq.~\eqref{eq:nonlinpie}.  With
\begin{equation}
	P\equiv P_\theta,
	\qquad
	\eta\equiv r^2C_\theta ,
\end{equation}
and using the definition
\begin{equation}
	F(P)=P\,\xi^2 ,
\end{equation}
Eq.~\eqref{eq:nonlinpie} can be written as
\begin{equation}
	\frac{dF}{dP}=G_{\rm NL}(P,F),
\end{equation}
where
\begin{equation}
	\boxed{
		G_{\rm NL}(P,F)
		=
		\frac{2F}{P}
		\mp
		\frac{4\eta}{P}\sqrt{\frac{F}{P}}
		-
		\frac{4P^2}{\hbar^2}
		+
		4\Lambda_\theta
		+
		\frac{\eta^2}{P^2}.
	}
	\label{eq:GNL}
\end{equation}
Two analytically reducible limiting equations are useful for estimating
the nonlinear solution.  For
\begin{equation}
	\Lambda_\theta=0,
	\qquad
	\frac{\eta^2}{P^2}\rightarrow0 ,
\end{equation}
one obtains
\begin{equation}
	\boxed{
		G_2(P,F)
		=
		\frac{2F}{P}
		\mp
		\frac{4\eta}{P}\sqrt{\frac{F}{P}}
		-
		\frac{4P^2}{\hbar^2}.
	}
	\label{eq:G2}
\end{equation}
In the complementary small-$P$ limit,
\begin{equation}
	\Lambda_\theta=0,
	\qquad
	{4P^2}\rightarrow0 ,
\end{equation}
the corresponding equation is
\begin{equation}
	\boxed{
		G_3(P,F)
		=
		\frac{2F}{P}
		\mp
		\frac{4\eta}{P}\sqrt{\frac{F}{P}}
		+
		\frac{\eta^2}{P^2}.
	}
	\label{eq:G3}
\end{equation}
The three differential flows satisfy
\begin{equation}
	G_{\rm NL}-G_2
	=
	4\Lambda_\theta+\frac{\eta^2}{P^2},
\end{equation}
whereas
\begin{equation}
	G_3-G_{\rm NL}
	=
	4\left(
	\frac{P^2}{\hbar^2}-\Lambda_\theta
	\right).
\end{equation}
Hence, in the region
\begin{equation}
	P^2\geq\hbar^2\Lambda_\theta ,
\end{equation}
one has the differential hierarchy
\begin{equation}
	\boxed{
		G_2(P,F)
		\leq
		G_{\rm NL}(P,F)
		\leq
		G_3(P,F).
	}
	\label{eq:Gbounds}
\end{equation}
Let $F_2$, $F_{\rm NL}$ and $F_3$ denote solutions generated from common
initial data by Eqs.~\eqref{eq:G2}, \eqref{eq:GNL} and \eqref{eq:G3},
respectively. On a common regular monotonic branch, the comparison
hierarchy implies the corresponding ordering of the reconstructed
momenta and angular actions, in accordance with the standard comparison
theorems for first-order nonlinear differential equations
\cite{LakshmikanthamLeela,Pachpatte}. A formal statement of the
comparison conditions and their preservation under the transformations
$F\rightarrow P_\theta\rightarrow S_\theta$ is given in \ref{AppendixB}.

\begin{equation}
	P_2(\theta)
	\lesssim
	P_{\rm NL}(\theta)
	\lesssim
	P_3(\theta),
\end{equation}
for the increasing branch; the ordering is reversed when the common
branch orientation is reversed.  Since
\begin{equation}
	\frac{dS_\theta}{d\theta}
	=
	P_\theta
	+
	\hbar\frac{\Phi}{\Phi_0},
\end{equation}
the same branchwise ordering is inherited by the corresponding angular
actions.  With a common choice $S_\theta(0)=0$,
\begin{equation}
	\boxed{
		S_2(\theta)
		\lesssim
		S_{\rm NL}(\theta)
		\lesssim
		S_3(\theta)
	}
	\label{eq:Sbounds}
\end{equation}
on the increasing branch and over the interval in which
$P^2\geq\hbar^2\Lambda_\theta$ and the three comparison solutions remain
regular.
\vspace{0.5pc}

\begin{figure}[!h]
	\centering
	\includegraphics[
	width=0.95\textwidth,
	height=.57\textheight,
	keepaspectratio
	]{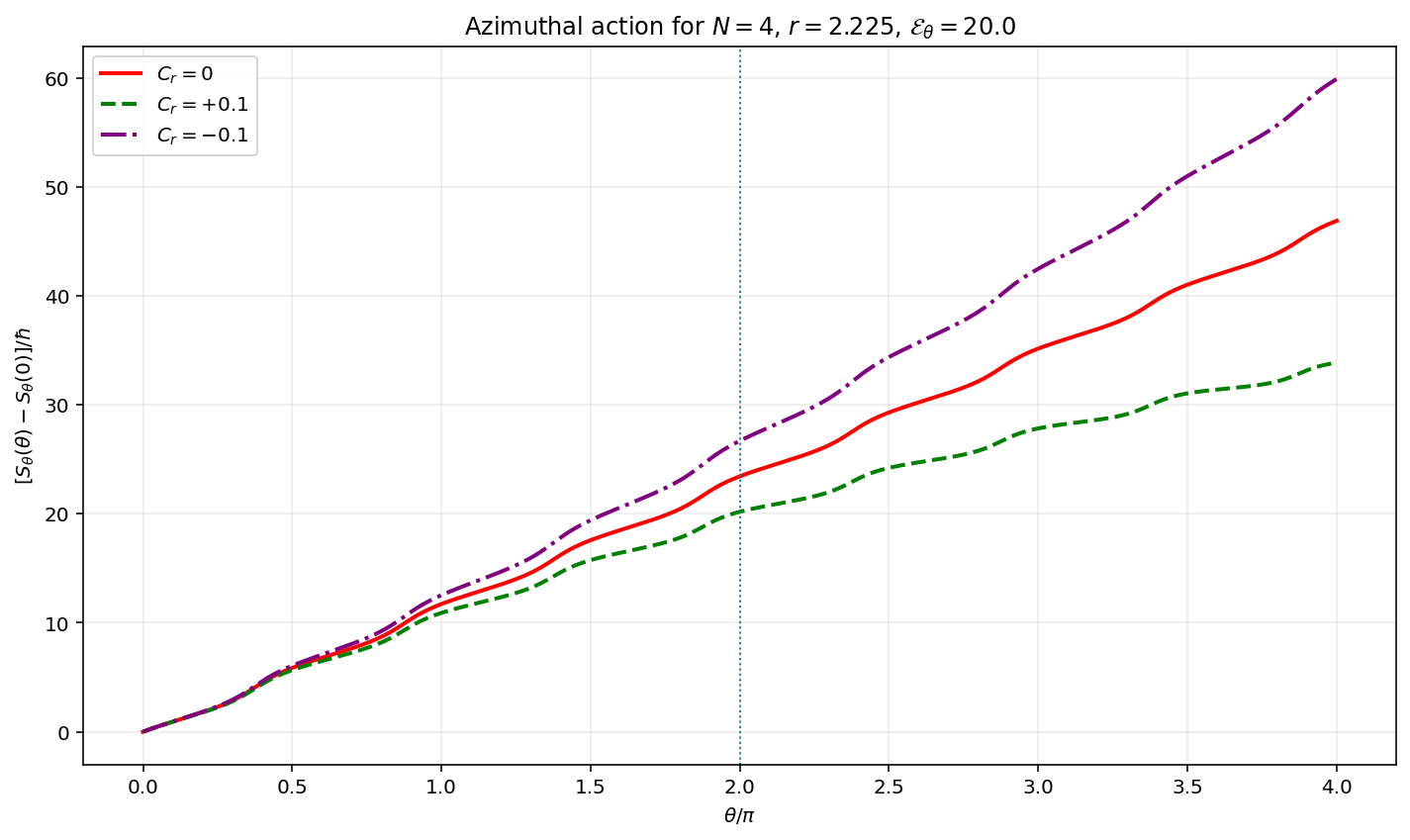}
	\caption{Azimuthal action \([S_\theta(\theta)-S_\theta(0)]/\hbar\) for the representative branch \(N=4\), \(r=2.225\), and \(\mathcal{E}_\theta=20.0\). The zero-current quasi-linear reference (\(C_r=0\), red) is compared with illustrative non-zero-current bounds (\(C_r=\pm0.1\)). The non-zero-current branches preserve the staircase-like trend of the zero-current solution while acquiring a weak quadratic curvature in \(\theta\); positive \(C_r\) produces downward bending, whereas negative \(C_r\) produces upward bending.}
	\label{fig:SthetaCrSec}
\end{figure}

\noindent
For the weak-current reduction $F_2$, the corresponding action has the
closed form
\begin{equation}
	\begin{split}
		S_2(\theta)
		={}
		-\frac{\hbar}{s}
		\left[
		\tan^{-1}\left\{
		\frac{\hbar}{2}
		\left(
		u_0-\frac{s\mathcal C_2}{2}\theta
		\right)
		\right\}
		-
		\tan^{-1}\left(
		\frac{\hbar u_0}{2}
		\right)
		\right]
		+
		\hbar\frac{\Phi}{\Phi_0}\theta ,
	\end{split}
\end{equation}
The current-dominated reduction $F_3$ admits two constant-slope
solutions,
\begin{align}
	S_{3a}(\theta)
	&=
	\left[
	P_\theta(0)+\hbar\frac{\Phi}{\Phi_0}
	\right]\theta
	-\frac{r^2C_r}{6}\theta^2 ,
	\\
	S_{3b}(\theta)
	&=
	\left[
	P_\theta(0)+\hbar\frac{\Phi}{\Phi_0}
	\right]\theta
	-\frac{r^2C_r}{2}\theta^2 .
\end{align}
The \(S_3(\theta)\) comparison bound is chosen appropriately between \(S_{3a}\) and \(S_{3b}\), depending on the sign of \(C_r\). The nonlinear bounds for \(\pm C_r\), relative to the zero-current reference, are illustrated in Figure~\ref{fig:SthetaCrSec}.

\paragraph{Remark:} The radial non-zero-current contribution is bounded independently by the monotonicity of the lower incomplete gamma function. Since the same current parameter \(C_r\) enters both sectors and is fixed by \(\Delta S_r+\Delta S_\theta=0\) together with \(C_\theta=-C_r\), a lower bound in one sector is accompanied by a compensating upper bound in the other under current conservation. For the angular comparison bounds, the magnetic and non-zero-current contributions also share the same radial scaling, since \(\Phi\propto Br^2\). Thus the \(F_3\) actions may be written schematically as \(S_3\sim r^2\theta(\mathcal B-\alpha C_r\theta)\), with \(\alpha=1/2\) or \(1/6\), showing that the current contribution acts as an angular distortion of the field-like coefficient proportional to \(C_r\theta\).
\section{Radial singularity, JWKB motivation, and minimum regularisation}
\label{sec:sec4}

The stationary-current construction and radial regularisation
address distinct aspects of the Landau problem. The latter
arises from the inverse-square term generated by cylindrical
separation, which makes $r=0$ a singular point of the radial
equation.
\vspace{0.5pc}

\noindent
The treatment of radial singularities is familiar from
semiclassical quantum mechanics. The Langer construction
modifies the treatment of the centrifugal term in WKB
theory~\cite{Langer}, while recent exact-WKB analyses
emphasise the contribution of the regular singular point
at the origin to angular-momentum-dependent
quantisation~\cite{ExactWKB}. These results motivate
examining the radial singularity directly within the
Bohm--Madelung amplitude--momentum formulation.
\vspace{0.5pc}

\noindent
The Fisher-information construction of Ref.~\cite{ASR}
provides the minimum canonical prescription
\[
x_i p_i=\frac{\hbar}{2}.
\]
Since the singularity under consideration occurs in
the radial equation, we apply this prescription
selectively:
\begin{equation}
	p_r=\frac{\hbar}{2r}.
	\label{eq:minimum-radial-momentum}
\end{equation}
The azimuthal and axial sectors are retained in their
native forms wherever their solutions remain regular.
\vspace{0.5pc}

\noindent
For the non-zero-current radial equation,
\[
\frac{1}{rR^2}
\frac{d}{dr}\left(rR^2p_r\right)=C_r,
\]
the minimum prescription gives
\begin{equation}
	R^2(r)\propto
	\exp\!\left(\frac{C_r r^2}{\hbar}\right).
	\label{eq:regularised-current-amplitude}
\end{equation}
This expression characterises the local amplitude
obtained from the prescribed radial momentum and
current equation. Its compatibility with the complete
stationary Hamilton--Jacobi equation is not assumed.
\vspace{0.5pc}

\noindent
The spectral consequence of the prescription can be
examined separately in the reduced radial equation.
The momentum contribution
\[
\frac{p_r^2}{2m}
=
\frac{\hbar^2}{8mr^2}
\]
adds an inverse-square term to the radial
Hamilton--Jacobi equation. With the conventional
angular separation eigenvalue retained, the resulting
radial amplitude equation takes the form
\begin{equation}
	R''+\frac{1}{r}R'
	+
	\left[
	\kappa_r^2-\beta^2r^2
	-\frac{l^2+\tfrac14}{r^2}
	\right]R=0.
	\label{eq:minimum-regularised-radial}
\end{equation}
Here $\kappa_r^2$ retains the same form of the Ermakov branch, Eq.~\eqref{eq:kappar}.

\vspace{0.5pc}

\noindent
In the unregularised equation, the regular
near-origin behaviour is $R(r)\sim r^{|l|}$.
The additional inverse-square contribution instead
gives
\begin{equation}
	R(r)\sim r^\nu,
	\qquad
	\nu=\sqrt{l^2+\frac14}.
	\label{eq:regularised-radial-index}
\end{equation}
The regularised radial solutions consequently retain
the associated-Laguerre structure, with a generally
noninteger index $\nu$. Unlike the conventional
Langer correction, the shift considered here follows
from the specified radial Bohm--Madelung
amplitude--momentum prescription. Its spectral
consequences are examined in the following section.

\section{Regularisation-induced spectral splitting}
\label{sec:sec5}

The regularised radial index
\[
\nu=\sqrt{l^2+\frac14}
\]
leads to the energy spectrum
\begin{equation}
	E_{n_rl}^{reg}(k_z)
	=
	\frac{\hbar\omega_c}{2}
	\left(
	2n_r+\sqrt{l^2+\frac14}+1
	\right)
	+
	\frac{\hbar^2k_z^2}{2m},
	\qquad
	\omega_c=\frac{|q|B}{m}.
\end{equation}
The noninteger radial index modifies the angular-index
dependence of the energy levels, resulting in a re-ordering
of the spectrum and lifting of the corresponding Landau
degeneracy. This spectral modification follows from the
minimum radial regularisation and is independent of the
nonlinear dynamics associated with non-zero stationary
currents.
\vspace{0.5pc}

\noindent
The contribution arising solely from the modified radial index can be isolated as

\[ 
\boxed{\Delta E_{\mathrm{rad}}(l)= \frac{\hbar\omega_c}{2} \left(\sqrt{l^2+\frac14}-|l|\right).}
 \]
This correction is positive and decreases monotonically with \(|l|\), attaining \(\hbar\omega_c/4\) at \(l=0\). For large \(|l|\), it approaches \(\hbar\omega_c/(16|l|)\). The radial-index modification is therefore most pronounced for small angular quantum numbers and becomes progressively weaker at higher \(|l|\).

\section{Summary and conclusion}

The stationary Bohm--Madelung formulation of the Landau problem reveals two structurally different effects which should not be conflated.
\vspace{0.5pc}

\noindent
The first concerns probability-amplitude dynamics. For \(C_i=0\), the separated system retains the Ermakov--Pinney reference structure. The zero-current azimuthal sector additionally admits periodic branches in which angular periodicity constrains the magnetic flux. When non-zero stationary currents are permitted, we restrict the globally conserved branch to
\[
C_z=0,\qquad C_\theta=-C_r.
\]
The radial sector remains analytically tractable, with its accumulated current contribution expressible through finite sums of lower incomplete gamma functions. The azimuthal sector is intrinsically nonlinear for \(C_\theta\neq0\). Two complementary solvable reductions provide comparison flows that bound the full nonlinear momentum and, on common regular monotonic branches, the corresponding angular action.
\vspace{0.5pc}

\noindent
The second effect concerns the energy spectrum. The radial separated equation contains an intrinsic inverse-square singularity, motivating comparison with the Langer and exact-WKB treatment of radial Schr\"{o}dinger problems. Applying the minimum Fisher-based prescription only to this singular radial sector gives
\[
\nu=\sqrt{l^2+\frac14},
\]
which modifies the radial Laguerre index and produces an explicit $l$-dependent contribution to the energy. The resulting lifting of the \(n_r,l\) degeneracy is therefore a regularisation-driven spectral mechanism, distinct from the stationary-current redistribution.
\vspace{0.5pc}

\noindent
The resulting picture is therefore one of differential ordering rather than global prescription: the Ermakov--Pinney structure provides the reference branch, stationary-current conservation generates nonlinear radial--azimuthal redistribution constrained by branchwise comparison bounds, and minimum radial regularisation controls the singular sector through a modified centrifugal index. The latter produces an explicit \(l\)-dependent spectral correction and lifts the corresponding Landau degeneracy. Together, these results separate current-driven nonlinear dynamics from regularisation-driven spectral restructuring while treating both within the same stationary Bohm--Madelung framework.
	
\paragraph*{Declaration of Interests\\}
\noindent The research is an independent work of the author, self-funded, and declares no conflicts of interest with the employer, IBM Research, Albany, NY, USA. While care has been taken to acknowledge relevant prior work, any inadvertent omissions are unintentional. The author welcomes appropriate references and perspectives to be incorporated in future revisions.

\paragraph*{Use of Generative AI \\}
\noindent Generative AI tools were used in the preparation of this manuscript solely for language editing, grammar refinement, and equation cross-verification. These tools were not used to generate the scientific results, derivations, or core interpretations presented here. All mathematical checks, technical conclusions, and final editorial decisions remain the sole responsibility of the author.

\appendix
\renewcommand{\thesection}{Appendix~\Alph{section}}
\renewcommand{\theequation}{\Alph{section}.\arabic{equation}}

\section{Bohm--Madelung reduction in cylindrical coordinates}
\label{AppendixA}
\setcounter{equation}{0}
\renewcommand{\thesection}{\Alph{section}}
\numberwithin{equation}{subsection}
\renewcommand{\theequation}
{\Alph{section}.\arabic{subsection}.\arabic{equation}}
\subsection{Stationary Bohm--Madelung formulation}
We consider a particle of mass $m$ and charge $q$ in a uniform magnetic field
$\bm B = B\hat{\bm z}$, with vanishing scalar potential. The stationary Schr\"{o}dinger
equation with minimal coupling reads ~\cite{Landau3}
\begin{equation}
	\frac{1}{2m}(\bm p - q\bm A)^2 \psi = E\psi,
\end{equation}
with the polar decomposition 
\begin{equation}
	\psi = \sqrt{\rho}\, e^{iS/\hbar}.
\end{equation}
where 

\[\rho(r,\theta,z) = R^2(r)\Theta^2(\theta)Z^2(z)\]

\subsection{Coordinate system and notation}

We work in cylindrical coordinates $(r,\theta,z)$ with metric factors
\begin{equation}
	ds^2 = dr^2 + r^2 d\theta^2 + dz^2,
	\qquad
	h_r = 1,\quad h_\theta = r,\quad h_z = 1.
\end{equation}

\noindent	The vector potential for a uniform magnetic field $B\hat{z}$ is taken in symmetric gauge as
\begin{equation}
	\mathbf{A} = A_\theta \,\hat{\theta} = \frac{Br}{2}\,\hat{\theta}.
\end{equation}

\noindent	We adopt the consistent Bohmian definition of momentum components as the \emph{physical gradient components} of the action:
\begin{equation}
	p_r = \partial_r S, \qquad
	p_\theta = \frac{1}{r}\partial_\theta S, \qquad
	p_z = \partial_z S.
\end{equation}
\noindent The stationary de Broglie--Bohm continuity equation is
\begin{equation}
	\label{eq:divfree}
	\nabla\cdot \mathbf J =0,
	\qquad
	\mathbf J=\frac{\rho}{m}\,(\nabla S-q\mathbf A).
\end{equation}
\subsection{Bohmian Hamiltonian with minimal coupling}

The stationary de Broglie--Bohm Hamiltonian is
\begin{equation}
	H_B = \frac{1}{2m}(\nabla S - q\mathbf{A})^2 + V_Q,
\end{equation}
where the quantum potential is
\begin{equation}
	V_Q = -\frac{\hbar^2}{2m}\frac{\nabla^2 \sqrt{\rho}}{\sqrt{\rho}}.
\end{equation}

\noindent Using the cylindrical form of $\nabla S$, we obtain
\begin{equation}
	\nabla S - q\mathbf{A}
	=
	\hat{r}\,p_r
	+
	\hat{\theta}\left(p_\theta - \frac{qBr}{2}\right)
	+
	\hat{z}\,p_z.
\end{equation}

\noindent Hence the Bohmian Hamiltonian becomes
\begin{equation}
	H_B
	=
	\frac{1}{2m}
	\left[
	p_r^2
	+
	\left(p_\theta - \frac{qBr}{2}\right)^2
	+
	p_z^2
	\right]
	+ V_Q.
\end{equation}

\noindent The stationary Bohm equation $H_B = E$ with expanded form of $V_Q$ becomes
\begin{equation}
	\label{eq:hamiltonianbohm}
	\frac{p_r^2}{2m}
	+
	\frac{p_\theta^2}{2m}
	-
	\frac{qBr}{2m}p_\theta
	+
	\frac{q^2 B^2 r^2}{8m}
	+
	\frac{p_z^2}{2m}
	-\frac{\hbar^2}{2m}
	\left(
	\frac{R''}{R}
	+
	\frac{1}{r}\frac{R'}{R}
	+
	\frac{1}{r^2}\frac{\Theta''}{\Theta}
	+
	\frac{Z''}{Z}
	\right)
	= E.
\end{equation}	

\noindent	The divergence condition Eq.~\eqref{eq:divfree} gives us the identity 
\begin{equation}
	\frac{1}{r}\frac{\partial}{\partial r}(r\rho p_r)
	+\frac{1}{r}\frac{\partial}{\partial \theta}
	\left[\rho\left(p_\theta-\frac{qBr}{2}\right)\right]
	+\frac{\partial}{\partial z}(\rho p_z)=0.
\end{equation}

\noindent We separate the above equation in component form such that

\[\frac{1}{r}\frac{d}{dr}(rR^2(r)p_r) = C_rR^2(r),\]
\[\frac{1}{r}\frac{d}{d\theta}
\left[\Theta^2(\theta)\left(p_\theta-\frac{qBr}{2}\right)\right]= C_\theta\Theta^2(\theta)\]
and
\[\frac{d}{dz}(Z^2(z)p_z) = C_zZ^2(z).\]

\noindent
with \[C_r+C_\theta+C_z =0.\]

\renewcommand{\thesection}{Appendix~\Alph{section}}
\section{Nonlinear comparison bounds for the non-zero-current angular equation}
\label{AppendixB}
\renewcommand{\thesection}{\Alph{section}}
\numberwithin{equation}{subsection}
\renewcommand{\theequation}
{\Alph{section}.\arabic{subsection}.\arabic{equation}}
\subsection{Nonlinear angular equation}
The angular equation in the presence of a non-zero current may be
written in terms of
\begin{equation}
	P\equiv P_\theta
	=
	\partial_\theta S
	-
	\hbar\frac{\Phi}{\Phi_0},
	\qquad
	\xi=\frac{P'}{P},
\end{equation}
where the prime denotes differentiation with respect to $\theta$.
Introducing
\begin{equation}
	\eta=r^2C_\theta=-r^2C_r,
\end{equation}
the angular equation takes the form
\begin{equation}
	\xi^2
	-2P\xi\frac{d\xi}{dP}
	-\frac{4\eta}{P}\xi
	-\frac{4P^2}{\hbar^2}
	+4\Lambda_\theta
	+\frac{\eta^2}{P^2}
	=0 .
	\label{eq:app_xi_full}
\end{equation}
This is equivalent to the nonlinear first-order equation
Eq.~\eqref{eq:nonlinpie} used in the main text.
\vspace{0.5pc}

\noindent
Defining
\begin{equation}
	F(P)=P\xi^2 ,
	\label{eq:app_Fdef}
\end{equation}
one obtains
\begin{equation}
	\frac{dF}{dP}
	=
	G_{\rm NL}(P,F),
\end{equation}
with
\begin{equation}
	\boxed{
		G_{\rm NL}(P,F)
		=
		\frac{2F}{P}
		\mp
		\frac{4\eta}{P}\sqrt{\frac{F}{P}}
		-
		\frac{4P^2}{\hbar^2}
		+
		4\Lambda_\theta
		+
		\frac{\eta^2}{P^2}.
	}
	\label{eq:app_GNL}
\end{equation}
The square-root term is the explicit nonlinear contribution associated
with the non-zero-current branch.

\subsection{Weak-current comparison solution $F_2$}

A first analytically controlled limit is obtained by setting
\begin{equation}
	\Lambda_\theta=0
\end{equation}
and neglecting the quadratic-current contribution
$\eta^2/P^2$, while retaining the term linear in $\eta$.  Equation
\eqref{eq:app_GNL} then reduces to
\begin{equation}
	\boxed{
		\frac{dF_2}{dP}
		=
		G_2(P,F_2)
		=
		\frac{2F_2}{P}
		\mp
		\frac{4\eta}{P}\sqrt{\frac{F_2}{P}}
		-
		\frac{4P^2}{\hbar^2}.
	}
	\label{eq:app_G2}
\end{equation}
For a weak current, $F_2$ may be expanded as
\begin{equation}
	F_2=F_{20}+\eta F_{21}+O(\eta^2).
\end{equation}
At leading order one finds
\begin{equation}
	F_{20}
	=
	P\xi^2,
\end{equation}
with the reduced equation admitting the first integral
\begin{equation}
	(P\xi)^2
	=
	\mathcal C_2 P^3
	-
	\frac{4P^4}{\hbar^2}.
	\label{eq:app_F20}
\end{equation}
Equivalently,
\begin{equation}
	P'
	=
	sP^{3/2}
	\sqrt{
		\mathcal C_2-\frac{4P}{\hbar^2}
	},
	\qquad s=\pm1 .
	\label{eq:app_P2prime}
\end{equation}
Introducing
\begin{equation}
	u^2
	=
	\frac{\mathcal C_2}{P}
	-
	\frac{4}{\hbar^2},
\end{equation}
Eq.~\eqref{eq:app_P2prime} reduces to
\begin{equation}
	\frac{du}{d\theta}
	=
	-\frac{s\mathcal C_2}{2},
\end{equation}
so that
\begin{equation}
	u(\theta)
	=
	u_0-\frac{s\mathcal C_2}{2}\theta .
\end{equation}
Consequently,
\begin{equation}
	P_2(\theta)
	=
	\frac{\mathcal C_2}{
		4/\hbar^2+
		\left(
		u_0-\frac{s\mathcal C_2}{2}\theta
		\right)^2
	} .
	\label{eq:app_P2}
\end{equation}
Since
\begin{equation}
	\frac{dS_\theta}{d\theta}
	=
	P_\theta
	+
	\hbar\frac{\Phi}{\Phi_0},
\end{equation}
and $S_\theta(0)=0$, integration of Eq.~\eqref{eq:app_P2}
gives
\begin{equation}
	\boxed{
		\begin{split}
			S_2(\theta)
			={}
			-\frac{\hbar}{s}
			\left[
			\tan^{-1}
			\left\{
			\frac{\hbar}{2}
			\left(
			u_0-\frac{s\mathcal C_2}{2}\theta
			\right)
			\right\}
			-
			\tan^{-1}
			\left(
			\frac{\hbar u_0}{2}
			\right)
			\right]
			+
			\hbar\frac{\Phi}{\Phi_0}\theta
			+O(C_r).
		\end{split}
	}
	\label{eq:app_S2}
\end{equation}
The constants $u_0$ and $\mathcal C_2$ are fixed by the initial data
of the reduced branch at \(\theta = 0\), with \(P_\theta(0)=P_0\).

\subsection{Small-$P$ comparison solution $F_3$}

A complementary current-dominated limit is obtained by taking
\begin{equation}
	\Lambda_\theta=0,
	\qquad
	{4P^2}\rightarrow0,
\end{equation}
while retaining the quadratic-current term.  The reduced equation is
\begin{equation}
	\boxed{
		\frac{dF_3}{dP}
		=
		G_3(P,F_3)
		=
		\frac{2F_3}{P}
		\mp
		\frac{4\eta}{P}\sqrt{\frac{F_3}{P}}
		+
		\frac{\eta^2}{P^2}.
	}
	\label{eq:app_G3}
\end{equation}
Using
\begin{equation}
	Z=P\xi \left(Z=\frac{dP}{d\theta}\right),
	\qquad
	Z^2=PF_3 ,
\end{equation}
the equation becomes
\begin{equation}
	2PZ\frac{dZ}{dP}
	=
	(3Z-\eta)(Z-\eta).
	\label{eq:app_Zeq}
\end{equation}
It is separable, and integration gives
\begin{equation}
	\boxed{
		\frac{(Z-\eta)^3}{3Z-\eta}
		=
		\mathcal C_3 P^3 .
	}
	\label{eq:app_F3implicit}
\end{equation}
where $\mathcal{C}_3$ is determined by the initial condition at \(\theta = 0\) with \(P_\theta(0)=P_0\).
\vspace{0.5pc}

\noindent
Equation~\eqref{eq:app_Zeq} also admits the constant-slope solutions
\begin{equation}
	Z=\eta,
	\qquad
	Z=\frac{\eta}{3}.
\end{equation}
Since $Z=P'=P\xi$, these give
\begin{equation}
	P_{3a}'=\eta,
	\qquad
	P_{3b}'=\frac{\eta}{3}.
\end{equation}
With $\eta=-r^2C_r$,
\begin{align}
	P_{3a}(\theta)
	&=
	P_\theta(0)-r^2C_r\theta,
	\\
	P_{3b}(\theta)
	&=
	P_\theta(0)-\frac{r^2C_r}{3}\theta .
\end{align}
The corresponding actions are
\begin{align}
	S_{3a}(\theta)
	&=
	\left[
	P_\theta(0)
	+
	\hbar\frac{\Phi}{\Phi_0}
	\right]\theta
	-
	\frac{r^2C_r}{2}\theta^2,
	\label{eq:app_S3a}
	\\
	S_{3b}(\theta)
	&=
	\left[
	P_\theta(0)
	+
	\hbar\frac{\Phi}{\Phi_0}
	\right]\theta
	-
	\frac{r^2C_r}{6}\theta^2.
	\label{eq:app_S3b}
\end{align}
The $S_3(\theta)$ bounds are appropriately chosen between
$S_{3a}$ and $S_{3b}$ depending on the sign of $C_r$.

\subsection{Preservation of the nonlinear hierarchy}

The three differential flows differ by
\begin{equation}
	G_{\rm NL}-G_2
	=
	4\Lambda_\theta+\frac{\eta^2}{P^2}
	\geq0 ,
	\label{eq:app_diff1}
\end{equation}
and
\begin{equation}
	G_3-G_{\rm NL}
	=
	4\left(
	\frac{P^2}{\hbar^2}
	-
	\Lambda_\theta
	\right).
	\label{eq:app_diff2}
\end{equation}
Consequently, over the region
\begin{equation}
	P^2\geq\hbar^2\Lambda_\theta,
	\label{eq:app_domain}
\end{equation}
one obtains
\begin{equation}
	\boxed{
		G_2(P,F)
		\leq
		G_{\rm NL}(P,F)
		\leq
		G_3(P,F).
	}
	\label{eq:app_Ghier}
\end{equation}
Let $F_2$, $F_{\rm NL}$ and $F_3$ be generated from common initial
data.  On a common regular interval with $P>0$ and $F>0$, the usual
comparison theorem for scalar first-order nonlinear differential
equations gives
\begin{equation}
	\boxed{
		F_2(P)
		\leq
		F_{\rm NL}(P)
		\leq
		F_3(P).
	}
	\label{eq:app_Fhier}
\end{equation}
The local uniqueness required for this construction follows from the
Picard--Lindel\"{o}f conditions away from the singular boundaries
$P=0$ and $F=0$.
From Eq.~\eqref{eq:app_Fdef},
\begin{equation}
	P'
	=
	\mu\sqrt{PF},
	\qquad
	\mu=\pm1 ,
	\label{eq:app_reconstruct}
\end{equation}
where $\mu$ is fixed on a monotonic branch.  For the increasing
branch, $\mu=+1$, the map $F\mapsto\sqrt{PF}$ is monotone for
$P>0$, and Eq.~\eqref{eq:app_Fhier} therefore gives
\begin{equation}
	P_2'
	\leq
	P_{\rm NL}'
	\leq
	P_3'.
\end{equation}
Applying the scalar comparison theorem once more, with common initial
momentum,
\begin{equation}
	P_2(0)
	=
	P_{\rm NL}(0)
	=
	P_3(0),
\end{equation}
gives
\begin{equation}
	\boxed{
		P_2(\theta)
		\leq
		P_{\rm NL}(\theta)
		\leq
		P_3(\theta).
	}
	\label{eq:app_Phier}
\end{equation}
For the oppositely oriented monotonic branch the inequalities are
reversed.
\vspace{0.5pc}

\noindent
The magnetic-flux contribution is common to all three solutions:
\begin{equation}
	S_j(\theta)
	=
	\int_0^\theta
	P_j(\vartheta)\,d\vartheta
	+
	\hbar\frac{\Phi}{\Phi_0}\theta,
	\qquad
	j=2,{ NL},3 .
\end{equation}
Since integration preserves ordering, Eq.~\eqref{eq:app_Phier}
implies
\begin{equation}
	\boxed{
		S_2(\theta)
		\leq
		S_{\rm NL}(\theta)
		\leq
		S_3(\theta)
	}
	\label{eq:app_Shier}
\end{equation}
on the common increasing branch, throughout the interval for which
Eq.~\eqref{eq:app_domain} and the regularity assumptions remain valid.
The ordering is reversed for the opposite monotonic orientation.
\vspace{0.5pc}

\noindent
Thus the analytically reducible $F_2$ and $F_3$ problems are not
components of a superposition of the nonlinear solution.  They provide
comparison solutions that locate the full non-zero-current angular
action within a controlled branchwise interval.  The result follows
from the standard theory of scalar differential inequalities and
comparison solutions \cite{LakshmikanthamLeela,Pachpatte}.
\end{document}